\documentstyle[epsfig,longtable]{aipproc}
\def\ga{\;\rlap{\lower 2.5pt
 \hbox{$\sim$}}\raise 1.5pt\hbox{$>$}\;}
\def\la{\;\rlap{\lower 2.5pt
   \hbox{$\sim$}}\raise 1.5pt\hbox{$<$}\;}
\def\gsim{\;\rlap{\lower 2.5pt
 \hbox{$\sim$}}\raise 1.5pt\hbox{$>$}\;}
\def\lsim{\;\rlap{\lower 2.5pt
   \hbox{$\sim$}}\raise 1.5pt\hbox{$<$}\;}
\def\lya{Ly$\alpha$}

\begin{document}
\begin{flushright}
{\footnotesize
FERMILAB-Conf-98/377-A}
\end{flushright}
\nopagebreak
\vspace{-\baselineskip}
\title{Empirical Constraints on the First Stars and Quasars}
 
\author{Z. Haiman$^{\dagger}$ and A. Loeb$^\star$}
\address{$^{\dagger}$Astrophysics Theory Group, Fermi National Accelerator
Laboratory, Batavia, IL 60510\\ 
$^{\star}$Harvard-Smithsonian Center for
Astrophysics, 60 Garden Street, Cambridge, MA 02138}

\lefthead{First Stars and Quasars}
\righthead{Haiman \& Loeb}
\maketitle

\begin{abstract}
Empirical studies of the first generation of stars and quasars in the Universe
will likely become feasible over the next decade.  The {\it Next Generation
Space Telescope} will provide direct imaging and photometry of sub--galactic
objects at $z\ga 10$, while microwave anisotropy experiments, such as MAP or
Planck, will set constraints on the ionization history of the intergalactic
medium due to these sources.  We describe the expected signals that will be
detectable with these future instruments.
\end{abstract}

\section*{Introduction}

The cosmic dark age~\cite{r96} ended when the first gas clouds condensed
out of the primordial fluctuations at redshifts $z=10-20$~\cite{l98}.
These condensations gave birth to the first star--clusters or mini--quasars
in the Universe. Previous observations have not yet probed this
epoch. Existing optical or infrared telescopes are capable of reaching out
to redshifts $z\la 6$, while current anisotropy experiments of the cosmic
microwave background (CMB) probe only the recombination epoch at $z\sim
10^3$. The remaining observational gap between these redshift regimes is
likely to be bridged over the coming decade. In particular, the planned
launch of the infrared Next Generation Space Telescope (NGST) will enable
direct imaging of sub--galactic objects at $z\ga 10$, while microwave
anisotropy satellites such as MAP or Planck will measure complementary
signatures from the reionization of the intergalactic medium at these
redshifts.

The detection~\cite{hcm98,d98} of Ly$\alpha$ emission from galaxies at
redshifts up to $z=5.7$ demonstrates that reionization due to the first
generation of sources must have occurred at yet higher redshifts;
otherwise, the damping wing of Ly$\alpha$ absorption by the neutral
intergalactic medium would have eliminated the Ly$\alpha$ line in the
observed spectrum of these sources~\cite{jm98,hs98}.  In this review, we
predict the redshift of reionization in popular cosmological models, and
estimate the signals from this epoch that will become detectable with the
NGST, MAP, or the Planck satellites.

\section*{Properties of the First Objects}

The first objects left behind a variety of fossil evidence of their
existence~\cite{cba84}, including the enrichment of the intergalactic
medium (IGM) with heavy elements, the reionization of the IGM, the
distortion of the CMB spectrum by dust-processed radiation~\cite{w81}, and
the production of stellar remnants.  In the following, we explore the
signatures of the early stars and mini-quasars using a simple
semi--analytical model, based on the Press--Schechter
formalism~\cite{ps74}.  We calibrate the total amount of light that stars
or mini--quasars produce based on data from redshifts $z\la 5$. The
efficiency of early star formation is calibrated based on the observed
metallicity of the intergalactic medium~\cite{t95,sc96}, while the early
quasars are constrained so as to match the quasar luminosity
function~\cite{p95} at redshifts $z\la 5$, as well as data from the Hubble
Deep Field (HDF) on faint point--sources.  We focus on a particular
cosmological model with a cosmological constant, ($\Omega_0,\Omega_\Lambda,
\Omega_{\rm b},h,\sigma_{8h^{-1}},n$)=(0.35, 0.65, 0.04, 0.65, 0.87, 0.96),
named the ``concordance model'' by Ostriker \& Steinhardt~\cite{os95}.  For
a discussion of other cosmological models, as well as a more detailed
description of our methods and results, we refer the reader to several
papers~\cite{hrl97,htl96,hl97a,hl97b,lh97,hl98a,hl98b,hml99}.  More
advanced 3--D numerical simulations have only now started to address the
complicated physics associated with the fragmentation, chemistry, and
radiative transfer of the primordial molecular clouds~\cite{abn98}, as well
as with the reionization of the IGM~\cite{go97}.

Following collapse, the gas in the first baryonic condensations is
virialized by a strong shock~\cite{b85}.  The shock--heated gas can only
continue to collapse and fragment if it cools on a timescale shorter than
the Hubble time.  In the metal--poor primordial gas, the only coolants that
satisfy this requirement~\cite{sz67} are neutral atomic hydrogen (H) and
molecular hydrogen (${\rm H_2}$). However, ${\rm H_2}$ molecules are
fragile, and are easily photo-dissociated throughout the universe by trace
amounts of starlight~\cite{sw67,hrl97} that are well below the level
required for complete reionization of the Universe.  Hence, most of the
sources that ionized the Universe formed inside objects with virial
temperatures $T_{\rm vir}\gsim10^4$K, or masses $\sim10^8~{\rm M_\odot}$,
which cooled via atomic transitions. Depending on the details of their
cooling and angular momentum transport, the gas in these objects fragmented
into stars, or formed a central black hole exhibiting quasar activity.
Although the first objects contained only a small fraction of the total
mass of the universe, they could have had a dramatic effect on the
subsequent evolution of the ionization and temperature of the rest of the
gas~\cite{dzn67}.  Since nuclear fusion releases $\sim 7$ MeV per baryon,
and accretion onto a black hole may release even more energy, and since the
ionization of a hydrogen atom requires only 13.6 eV, it is sufficient to
convert a small fraction of the baryonic mass into either stars or black
holes in order to ionize the rest of the Universe.

The cooling gas clouds eventually fragment into stars~\cite{cr86}.
Although the actual fragmentation process is likely to be rather complex,
the average fraction $f_{\rm star}$ of the collapsed gas converted into
stars can be calibrated empirically so as to reproduce the average
metallicity observed in the Universe at $z\approx 3$.  For the purpose of
this calibration, we use the average C/H ratio, inferred from CIV absorption
lines in \lya~forest clouds~\cite{t95,sc96}.  The observed ratio is between
$10^{-3}$ and $10^{-2}$ of the solar value~\cite{s97}.  If the carbon
produced in the early star clusters is uniformly mixed with the rest of the
baryons in the Universe, this implies $f_{\rm star}\approx$2--20\% for a
Scalo~\cite{s86} stellar mass function.  This number assumes inefficient
hot bottom burning, i.e. maximal carbon yields~\cite{rv81}, and includes a
factor of $\sim$3 due to the finite time required to produce carbon inside
the stars (in a Press--Schechter star formation history, only a third of
the total stellar carbon yield is produced and ejected by $z=3$).
Ultimately, 3-D simulations of the first generation of stars might be used
to infer the expected star--formation efficiency in the first generation of
gas clouds. Preliminary runs~\cite{n98} imply that $\sim 1\%$ of the gas
condenses into dense cores which could yield massive stars.

An even smaller fraction of the cooling gas might condense at the center of
the potential well of each cloud and form a massive black hole, exhibiting
mini--quasar activity.  In the simplest scenario we postulate that the peak
luminosity of each black hole is proportional to its mass, and there exists
a universal quasar light--curve in Eddington units.  This hypothesis is
motivated by the fact that for a sufficiently high fueling rate, quasars
are likely to shine at their maximum possible luminosity, which is some
constant fraction of the Eddington limit, for a time which is dictated by
their final mass and radiative efficiency.  Allowing the final black hole
mass $M_{\rm bh}$ to be a fixed fraction of the total halo mass $M_{\rm
halo}$, we find that there exists a universal light curve [$L(t)=L_{\rm
Edd}\exp(-t/t_0)$, with $t_0\sim10^{6}$ yr], for which the Press--Schechter
theory provides an excellent fit to the observed evolution of the
luminosity function (LF) of bright quasars between redshifts $2.6<z<4.5$.
The required black hole to halo gas mass ratio is $M_{\rm bh}/M_{\rm
gas}=10^{-3.2}\Omega_{0}/\Omega_{\rm b}=5.5\times10^{-3}$, close to the
typical value of $\sim6\times10^{-3}$ found for the ratio of black hole
mass to spheroid mass in a dozen nearby galaxies~\cite{k97,m98}. The
existence of massive black holes in the centers of low--mass galaxies such
as M32 or NGC 4486B~\cite{k97,m98} implies that the process of black hole
formation does not discriminate against galaxies of this type. Since
galaxies at $z\sim 10$ have a similar mass and velocity dispersion as these
low--redshift examples, it is conceivable that low--luminosity quasars
contributed significantly to the reionization of the Universe.

\begin{figure} 
\centerline{\epsfig{file=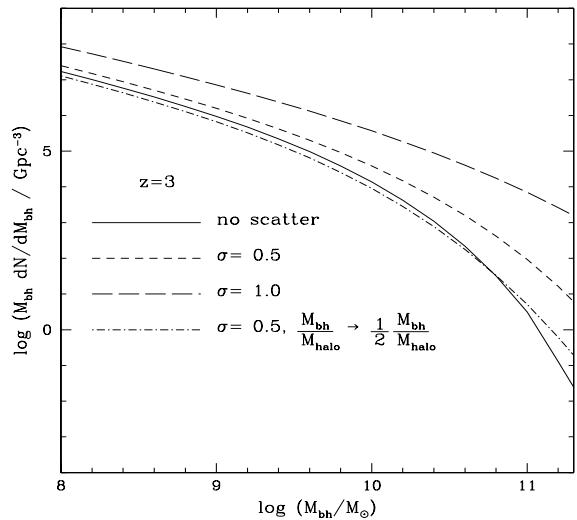,width=3.2in,height=2.9in}}
\vspace{10pt}
\caption{ {\it The comoving mass function of black holes at redshift $z=3$ when
a scatter is introduced to the logarithm of the ratio between the black hole
and halo masses, $\log (M_{\rm bh}/M_{\rm halo})$.}}\label{fig:scat}
\end{figure}

One does expect, however, that the ratio $M_{\rm bh}/M_{\rm gas}$ would
have a substantial intrinsic scatter.  Observationally, the scatter around
the average value of $\log (M_{\rm bh}/L)$ is 0.3~\cite{m98}, while the
standard deviation in $\log M_{\rm bh}/M_{\rm gas}$ has been found to be
$\sigma\sim 0.5$~\cite{vdm98}.  Such an intrinsic scatter would flatten the
predicted quasar LF at the high mass end, where the LF is a steep function
of black hole mass.  As an illustrative example, we show in
Figure~\ref{fig:scat} the mass function of black holes in the
Press--Schechter model of halos, with or without this scatter (solid versus
short-dashed lines).  In order to eliminate the flattening introduced by
the scatter, we find that the average black hole to halo mass ratio must be
reduced by $\sim 50\%$.  The dot--dashed line in Figure~\ref{fig:scat}
demonstrates that such a reduction would indeed compensate for the effect
of an intrinsic scatter in the relevant mass range ($10^{8}~{\rm
M_\odot}\lsim M_{\rm bh}\lsim 10^{10}~{\rm M_\odot}$).
Figure~\ref{fig:scat} also shows the effect of a more significant intrinsic
scatter ($\sigma\sim 1$) on the black hole mass function (long-dashed
lines).  We find that the predicted black hole mass function in the
presence of such a large scatter would be significantly different from any
model with a constant value for $M_{\rm bh}/M_{\rm gas}$.

In reality, the relation between the black hole and halo masses may be more
complicated than linear in reality.  With the introduction of additional
free parameters, a non--linear (mass and redshift dependent) relation
between the black--hole and halo masses can also lead to acceptable
fits~\cite{hnr98} of the observed quasar LF.  The nonlinearity in the
relation must be related to the physics of the formation process of
low--luminosity quasars, which was discussed in several
papers~\cite{hnr98,el95,l97}.  If the black hole formation efficiency
decreases in smaller halos, this would flatten the faint end of the LF, and
therefore could not compensate the effect of a large intrinsic scatter of
the type shown in Figure~\ref{fig:scat}.  Indeed, in order to fit the
bright end of the LF in a model with a large intrinsic scatter, one must
postulate that the black hole formation efficiency decreases in larger
halos.

\noindent

\noindent
\begin{figure} 
\centerline{\epsfig{file=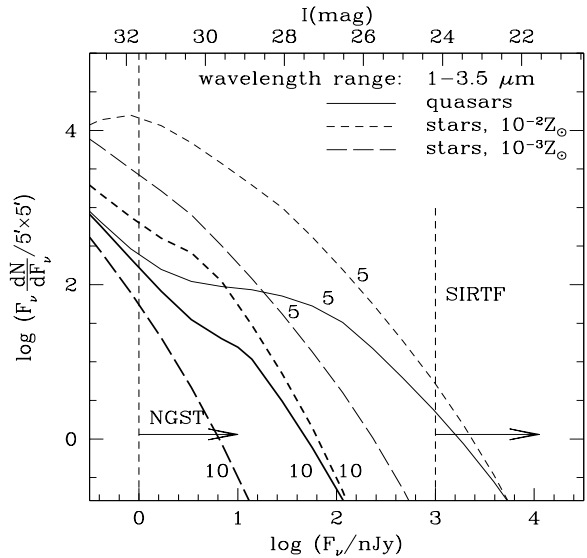,width=3.2in,height=2.9in}}
\vspace{10pt}
\caption{ {\it Infrared Number Counts. The solid curves refer to quasars,
while the long/short dashed curves correspond to star clusters with
low/high normalization for the star formation efficiency.  The curves
labeled ``5'' or ``10'' show the cumulative number of objects with
redshifts above $z=5$ or 10.}  }\label{fig:ncounts}
\end{figure}

\section*{Infrared Number Counts}

The Next Generation Space Telescope ({\it NGST}, \cite{ngst}) will be able to
detect the early population of star clusters and mini--quasars.  {\it NGST} is
scheduled for launch in 2007, and is expected to reach an imaging sensitivity
of $\sim 1$ nJy (S/N=10 at spectral resolution $\lambda/\Delta\lambda=3$) for
extended sources after several hours of integration in the wavelength range of
1--3.5$\mu$m. Figure~\ref{fig:ncounts} shows the predicted number counts in the
models described above, normalized to a $5^{\prime}\times5^{\prime}$ field of
view.  This figure shows separately the number per logarithmic flux interval of
all objects with redshifts $z>5$ (thin lines), and $z>10$ (thick lines).  The
number of detectable sources is high; {\it NGST} will be able to probe about
$\sim100$ quasars at $z>10$, and $\sim200$ quasars at $z>5$ per field of view.
The bright--end tail of the number counts approximately follows a power law,
with $dN/dF_\nu\propto F_\nu^{-2.5}$.  The dashed lines show the corresponding
number counts of ``star--clusters'', assuming that each halo shines due to a
starburst that converts a fraction of 2\% (long--dashed) or 20\%
(short--dashed) of the gas into stars.  These lines indicate that {\it NGST}
would detect $\sim40-300$ star--clusters at $z>10$ per field of view, and
$\sim600-10^4$ clusters at $z>5$.  Unlike quasars, star clusters could in
principle be resolved if they extend over a scale comparable to the virial
radius of their dark matter halos~\cite{hl97b}.  The supernovae and
$\gamma$-ray bursts in these star clusters might outshine their hosts and may
also be directly observable~\cite{mr97,wl98}.

\begin{figure} 
\centerline{\epsfig{file=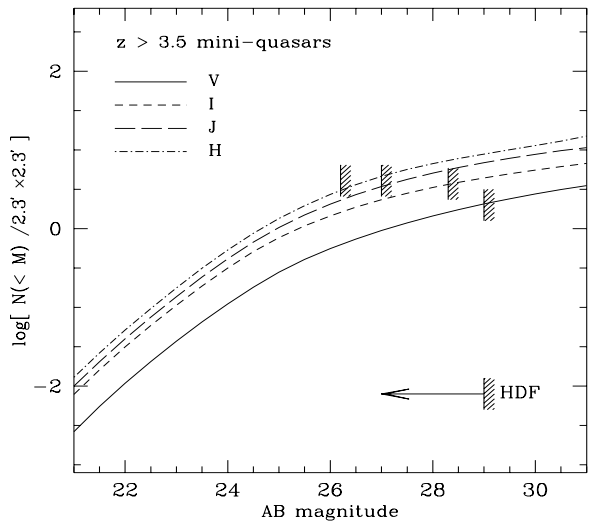,width=3.2in,height=2.9in}}
\vspace{10pt}
\caption{ {\it $V$, $I$, $J$, and $H$ counts for mini--quasars in a model with
a minimum halo circular velocity of $v_{\rm circ}= 75~{\rm km~s^{-1}}$.  This
model is consistent with HDF data in the $V$ band at the 5\% level.
Sensitivities are shown in each band for the same signal--to--noise ratio and
exposure time as the optical HDF.}}
\label{fig:nicmos}
\end{figure}

\section*{Constraints from the Hubble Deep Field}

High resolution, deep imaging surveys can be used to set important
constraints on semi--analytical models of the type described above.  The
properties of faint {\it extended} sources found in the Hubble Deep Field
(HDF)~\cite{m96} agree with detailed semi--analytic models of galaxy
formation~\cite{b98}.  On the other hand, the HDF has revealed only a
handful of faint {\it unresolved} sources, but none with the colors
expected for high redshift quasars~\cite{c99}.  The simplest mini--quasar
model described above predicts the existence of $\sim10$ B--band
``dropouts'' in the HDF, inconsistently with the lack of detection of such
dropouts up to the $\sim50\%$ completeness limit at $V\approx 29$ in the
HDF.  To reconcile the models with the data, a mechanism is needed for
suppressing the formation of quasars in halos with circular velocities
$v_{\rm circ} \lsim 50-75~{\rm km~s^{-1}}$.  This suppression naturally
arises due to the photo-ionization heating of the intergalactic gas by the
UV background after reionization~\cite{tw96,ns97}.  Alternative effects
could help reduce the quasar number counts, such as a change in the
background cosmology, a shift in the ``big blue bump'' component of the
quasar spectrum to higher energies due to the lower black hole masses in
mini--quasars, or a nonlinear black hole to halo mass relation; however,
these effects are too small to account for the lack of detections in the
HDF~\cite{hml99}.

The mini-quasars may not necessarily appear as point sources in the HDF if
their extended host galaxies are actually resolved by {\it HST}.  In fact,
twelve candidate sources of activity in the nuclei of galaxies at
high--redshifts ($z > 3.5$) have recently been identified~\cite{j98} in the
HDF.  All of these point--like sources are embedded in extended host
galaxies which are relatively bright ($V\sim 26-27$) and outshine their
AGNs by typically 1 mag. As a consequence, these AGNs would have been
missed by previous searches for isolated point--like sources~\cite{c99}.
In the models described above, the mini--quasars peak at a flux $\sim
1-3.5$ mag brighter in $V$ than their host galaxy, which is assumed to
undergo a starburst inside the same halo~\cite{hl98a}.  The twelve HDF
candidates must reflect faint AGN activity in bright galaxies of relatively
massive halos (analogous to a weak Seyfert activity), rather than faint AGN
activity in small halos as expected in our model.  The observed sources
might imply a phase in the history of massive halos that corresponds to an
additional low--luminosity tail of the quasar lightcurve discussed above.
An extended lightcurve would still be consistent with the luminosity
function derived from the bright AGN phase and could explain the existence
of the embedded AGNs in the HDF.  The detection of faint embedded AGNs
could also be related to the intrinsic scatter in the distribution of
$M_{\rm bh}/M_{\rm halo}$ relation, and reflect objects with unusually
small values of this mass ratio.

The longer infrared wavelengths, such as the $J$ and $H$ infrared bands, are
better suited for studying the Universe at $z\gsim5$.  Forthcoming data on
point--sources from NICMOS observations of the HDF\cite{t98} could improve the
constraints on mini-quasar models. In Figure~\ref{fig:nicmos}, we show the
expected number counts of mini--quasars in the $V$, $I$, $J$, and $H$ bands for
the model which is consistent with the optical HDF data. The number of objects
predicted in the $I$, $J$, and $H$ bands is higher than in the optical HDF.
The NICMOS data in these bands would either reveal several high redshift
mini--quasars or else place tighter constraints on quasar models than currently
possible using $V$ and $I$ data.  With the post--reionization feedback on halos
imposed by the optical HDF data, $v_{\rm circ}\geq 75~{\rm km~s^{-1}}$, we
still expect at least $\sim 5$ mini-quasars to be found at $z>3.5$ in the
NICMOS $J$ and $H$ bands.  A non--detection by NICMOS would translate to a
minimum circular velocity of $v_{\rm circ}\gsim 100~{\rm km~s^{-1}}$, or a
factor of $\sim 2$ increase in the low--mass cutoff for halos harboring
quasars.

\begin{figure} 
\centerline{\epsfig{file=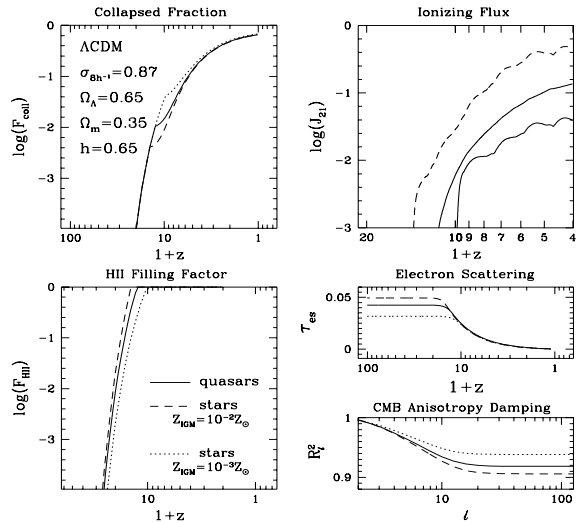,width=3.2in,height=2.9in}}
\vspace{10pt}
\caption{ {\it Reionization history.  Clockwise, the different panels show:
(i) the collapsed fraction of baryons; (ii) the background flux at the
Lyman limit; (iii) the volume filling factor of ionized hydrogen; and (iv)
the optical depth to electron scattering, and the corresponding damping
factor for the power--spectrum decomposition of microwave anisotropies as a
function of the spherical harmonic index $\ell$.  The solid curves refer to
quasars, while the dotted or dashed curves correspond to stars with a low
(2\%) or high (20\%) normalization for the star formation efficiency.}}
\label{fig:panels}
\end{figure}

\section*{Was the Universe Reionized by Stars or Mini--Quasars?}

Given either the star--formation or quasar black--hole formation histories,
we derive the reionization history of the IGM by following the radius of
the expanding Str\"omgren sphere around each source.  The reionization
history depends on the time--dependent production rate of ionizing photons,
their escape fraction, and the recombination rate of the IGM, which are all
functions of redshift.  The production rate of ionizing photons per quasar
follows from the median quasar spectrum~\cite{e94} and the light--curve we
derived.  The analogous rate per star follows from the time--dependent
composite stellar spectrum, constructed from standard stellar atmosphere
atlases~\cite{k93} and evolutionary tracks~\cite{s92}.  We computed the
escape fraction of ionizing photons in each halo, assuming ionization
equilibrium inside an isothermal sphere where both the stars and the gas
are distributed with a $1/r^2$ profile.  For quasars, we assumed that the
escape fraction is 100\%.  During reionization, we also assumed that the
formation of low--mass halos is suppressed by photo--ionization heating
inside the already ionized cosmological HII regions.

Figure~\ref{fig:panels} summarizes the resulting reionization histories
from stars or quasars in our models with a $\Lambda$CDM cosmology. The
results for stars are shown in two cases, one with $Z_{\rm IGM}=10^{-2}{\rm
Z_\odot}$ (dashed lines) and the other with $Z_{\rm IGM}=10^{-3}{\rm
Z_\odot}$ (dotted lines), to bracket the allowed IGM metallicity range.
The panels in Figure~\ref{fig:panels} show (clockwise) the total collapsed
fraction of baryons available for star or quasar formation; the evolution
of the average comoving flux, $J_{21}$ at the local Lyman limit frequency,
in units of $10^{-21}~{\rm erg~s^{-1}~cm^{-2}~Hz^{-1}~sr^{-1}}$; the
resulting evolution of the ionized fraction of hydrogen, $F_{\rm HII}$; and
the consequent damping of the CMB anisotropies. In the first two panels, we
imposed $v_{\rm circ}\geq 75~{\rm km~s^{-1}}$ at $z<z_{\rm reion}$. The
dashed and dotted curves indicate that stars ionize the IGM by a redshift
$9\lsim z\lsim13$; while the solid curve shows that quasars reionize the
IGM at $z\approx 11$.  This result can be understood in terms of the total
number of ionizing photons produced per unit halo mass; given our
normalizations of the efficiencies of star and quasar black hole formation,
the relative ratios of this number in the three cases are
$1\div0.37\div0.1$, respectively. A comparison of our quasar and stellar
template spectra shows that stars will not reionize HeII, while quasars
reionize HeII at essentially the H reionization redshift.  Therefore,
recent claims that HeII reionization might have been observed at $z\sim
3$~\cite{r97} could rule-out the presence of mini-quasars with hard spectra
extending to X--rays at high redshifts, if these claims are verified by
future observations.

\begin{figure} 
\centerline{\epsfig{file=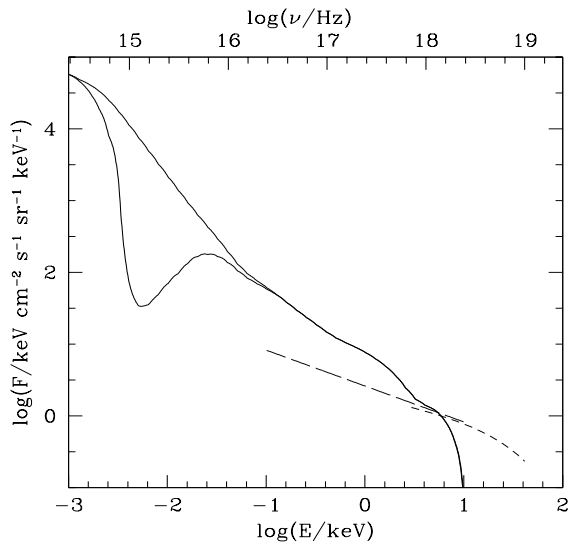,width=3.2in,height=2.9in}}
\vspace{10pt}
\caption{{\it Spectrum of the UV/X--ray background in the mini--quasar model,
assuming that the median X-ray spectrum of quasars~\protect\cite{e94} is
universal.  The solid curves show the spectrum with and without absorption by
the \lya~forest at high--redshifts. The short and long dashed lines show the
unresolved fraction (assumed to be 25\%) of the observed X--ray background
spectrum (from \protect\cite{xrb} and \protect\cite{fb92}).}}\label{fig:xrb}
\end{figure}

The X--ray background (XRB)~\cite{xrb,fb92} might provide another useful
constraint on the mini--quasar models.  In Figure~\ref{fig:xrb} we show the
predicted spectrum of the UV to the soft XRB at $z=0$ in these models
(solid lines).  In computing the spectrum, we included absorption by
neutral H and He in the IGM at $z>z_{\rm reion}=11$ and hydrogen
absorption~\cite{mad96} by the (extrapolated) Ly$\alpha$ forest at $z\leq
11$.  Also shown in this figure is the unresolved 25\% fraction of the
observed soft XRB\cite{xrb,fb92}.  The dashed lines in figure~\ref{fig:xrb}
represent an upper limit on any component of the XRB that could arise from
high--redshift quasar activity. As the figure shows, the mini--quasar
models overpredict the {\it unresolved flux} by a factor of $\sim 2-7$ in
the 0.1-1 keV range, as they produce a flux comparable to the entire soft
XRB flux.  If an even larger fraction of the XRB will be resolved into
low--redshift AGNs in the future, then the XRB could be used to place
stringent constraints on the X-ray spectrum or the abundance of the
mini--quasars discussed here.

\section*{Can The Reionization Redshift be Inferred from a Source
Spectrum?}

The spectrum of a source at a redshift $z_{\rm s}>z_{\rm reion}$ should
show a Gunn--Peterson (GP)~\cite{gp65} trough due to absorption by the
neutral IGM at wavelengths shorter than the local \lya~resonance at the
source, $\lambda_{\rm obs}<\lambda_\alpha(1+z_{\rm s})$. By itself, the
detection of such a trough would not uniquely establish the fact that the
source is located beyond $z_{\rm reion}$, since the lack of any observed
flux could be equally caused by: (i) ionized regions with some residual
neutral fraction, (ii) individual damped \lya~absorbers, or (iii) line
blanketing from lower column density \lya~forest absorbers.  On the other
hand, for a source located at a redshift $z_{\rm s}$ beyond but close to
reionization, $(1+z_{\rm reion}) < (1+z_{\rm s}) < \frac{32}{27} (1+z_{\rm
reion})$, the GP trough splits into disjoint Lyman $\alpha$, $\beta$, and
possibly higher Lyman series troughs, with some transmitted flux in between
these troughs. Although the transmitted flux is suppressed considerably by
the dense Ly$\alpha$ forest after reionization, it is still detectable for
sufficiently bright sources, and can be used to infer the reionization
redshift.

\begin{figure} 
\centerline{\epsfig{file=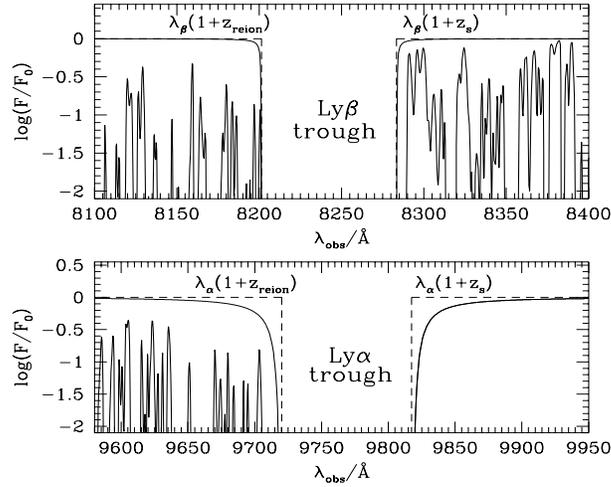,width=3.2in,height=2.9in}}
\vspace{10pt}
\caption{{\it Spectrum of a source at $z_{\rm s}=7.08$, assuming sudden
reionization at a redshift $z_{\rm reion}=7$.  The solid curves show the
spectrum without absorption by the high--redshift \lya~forest, and the dashed
lines show the spectrum when the damping wings are also ignored.}
}\label{fig:spect}
\end{figure}

As an example, we show in Figure~\ref{fig:spect} the simulated spectrum
around the Lyman $\alpha$ and $\beta$ GP troughs of a source at redshift
$z_{\rm s}=7.08$, assuming that reionization occurs suddenly at $z_{\rm
reion}=7$.  We have included the extrapolated effects of Ly$\alpha$
absorbers along the lines of sight, whose statistics were chosen so as to
obey the redshift dependence and absorption line characteristics of
observational data at $z<4.3$~\cite{fgs98}.  Although the continuum flux is
strongly suppressed, the spectrum contains numerous transmission features;
these features are typically a few \AA~wide, have a central intensity of a
few percent of the underlying continuum, and are separated by $\sim 10$\AA.
For sudden reionization, the nominal integration time of about 10 hours for
a $\sim10$ nJy sensitivity (with $\lambda/\Delta\lambda$=100 and S/N=10)
would be sufficient to determine $z_{\rm reion}$ up to a redshift of
$\sim7$ with a high precision, by determining the location of the
short--wavelength edge of the troughs shown in Figure~\ref{fig:spect}. More
gradual reionization would smear the edge of the GP trough. Based on the
extrapolation of the Ly$\alpha$ forest to higher redshifts, the required
sensitivity needs to be one or two orders of magnitude higher if $z_{\rm
reion}\sim8$ or $\sim9$ (see~\cite{hl98b} for details).

\begin{figure} 
\centerline{\epsfig{file=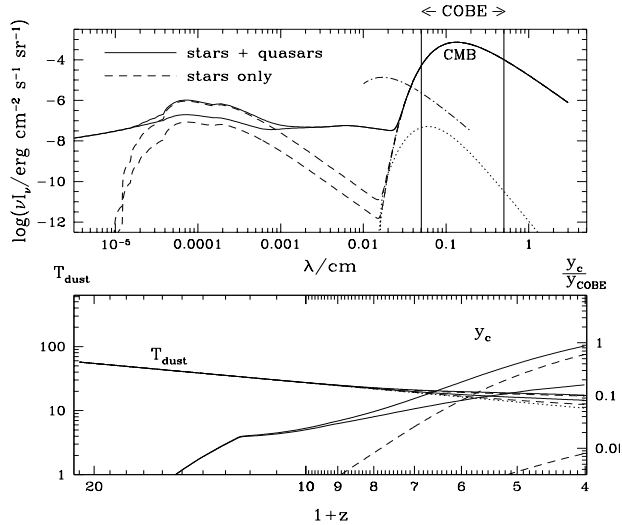,width=3.2in,height=2.9in}}
\vspace{10pt}
\caption{ {\it Effect of dust on the background flux.  The top panel shows the
comoving spectra in four different models at $z=3$, and the bottom panel shows
the corresponding evolution of the dust temperature and Compton y-parameter
(see text). }}\label{fig:dust}
\end{figure}

\section*{Signatures Imprinted on the CMB}

The free electrons produced by the reionization of the intergalactic medium
partially smooth--out the temperature anisotropies of the CMB via Thomson
scattering.  Given the ionized fraction of hydrogen as a function of redshift,
one can readily derive the electron scattering optical depth ($\tau_{\rm es}$),
as well as the anisotropy damping factor ($R^2_\ell$), as functions of the
spherical harmonic index $\ell$ of the multipole expansion of the anisotropies
on the sky~\cite{hw97}. As illustrated in the lower right panel of
Figure~\ref{fig:panels}, the amplitude of the anisotropies is reduced by
$\sim6-10\%$ on scales below the angular scale of the horizon at reionization
($\ell\ga 10$). Although small, this reduction is within the proposed
sensitivities of the future MAP and Planck satellites, provided that both
temperature and polarization anisotropy data will be gathered in these
experiments (see Table 2 in~\cite{zss97}).

In addition, the dust that is inevitably produced by the first type II
supernovae, absorbs the UV emission from early stars and quasars and re-emits
this energy at longer wavelengths, where it distorts the CMB spectrum.  We have
calculated this spectral distortion assuming that each type II supernova yields
${\rm 0.3M_\odot}$ of dust with the wavelength-dependent opacity of Galactic
dust~\cite{m90}. We have conservatively assumed that similarly to the observed
intergalactic mixing of metals, this dust gets uniformly distributed throughout
the intergalactic medium.  Clumpiness of the dust around UV sources would only
enhance our predicted spectral distortion.  The top panel of
Figure~\ref{fig:dust} shows the resulting total comoving spectrum of the
radiation background (CMB + direct quasar and/or stellar emission + dust
emission) at $z=3$. More distortion could be added between $0<z<3$ by dust and
radiation from galaxies.  For reference, we also show by the dot--dashed lines
the recently detected cosmic infrared background (CIB,~\cite{cib}), and the
typical dust peak in our calculations (dotted lines).

Figure~\ref{fig:dust} shows that in our models the dust emission peaks at a
wavelength which is an order of magnitude longer than that of the CIB peak.
This is a result of our assumption of a homogeneous dust distribution, and the
consequent cold dust temperature. An inhomogeneous distribution of dust would
raise its temperature, and could contribute significantly to the observed CIB
peak.  The deviation from the pure 2.728$(1+z)$K blackbody shape is quantified
by the Compton $y$--parameter, whose redshift evolution is shown in the bottom
panel. Ignoring the UV flux from quasars, we obtain $1.1\times 10^{-7}<y_{\rm
c}<8.2\times 10^{-6}$ at $z=3$ (dashed lines), just below the upper
limit~\cite{f96} set by COBE, $y<1.5\times10^{-5}$.  Adding the UV flux of
quasars increases the $y$--parameter to $4.1\times10^{-6}<y_{\rm
c}<2\times10^{-5}$.  The distortion by the intergalactic dust may have been
overestimated in our calculation as we ignored the absorption of the UV
background by the neutral component of the IGM. Nevertheless, a substantial
fraction ($\sim10$--$50$\%) of the total $y$--parameter results simply from the
direct far-infrared emission by early quasars and could be present even in the
absence of any intergalactic dust.

\section*{Acknowledgements}

We are grateful to Martin Rees for sharing his insights during our
collaboration on some of the topics described in this review.  ZH
acknowledges support at Fermilab by the DOE and the NASA grant NAG 5-7092.

\end{document}